\documentstyle[twocolumn,aps]{revtex}

\begin{document}
\draft

\twocolumn[\hsize\textwidth\columnwidth\hsize\csname @twocolumnfalse\endcsname

\title{Multimode electron-phonon coupling \\
in the two-dimensional $t-J-$Holstein model''}

\author{D. Poilblanc$^{1,2}$, T. Sakai$^1$, D. J. Scalapino$^{2}$ and W. Hanke$^{3,2}$}
\address{
$^{1}$Laboratoire de Physique Quantique, Universit\'e Paul Sabatier,
31062 Toulouse, France\\
$^{2}$Department of Physics, University of California Santa Barbara, CA 93106\\
$^{3}$Institute for Physics, University of Wurzburg, D-8700 Wurzburg, Germany\\
}

\date{October 95}
\maketitle 

\begin{abstract}
Polaronic effects are investigated in the two dimensional t--J model 
(describing
Cu spins and holes of the planes in high-$T_c$ cuprates) in the presence 
of in-plane dynamic oxygen breathing modes.
We focus on the effect of dynamic multi-phonon modes. 
Spin and charge structure factors are computed.  
We found evidence for self-localisation
of the holes and {\it increased} short range antiferromagnetic 
spin-spin correlations between the remaining spins.
This effect persists when all the in-plane oxygen vibration 
degrees of freedom are taken into account. 
This suggests that dynamic phonons could play a key role in stabilizing
inhomogeneous stripe phases or domain wall structures.
\end{abstract}

\vskip2pc]
\narrowtext

The interplay of the electron-phonon interaction and localization,
charge density wave (CDW) and spin density wave (SDW) correlations
in strongly interacting electron systems is an interesting issue. 
In the static Born-Oppenheimer $m\rightarrow\infty$ limit, 
Zhong and Sch\"uttler \cite{static} studied Hubbard and t--J models
with an onsite Holstein electron-phonon apical oxygen coupling. They
found on Hubbard ($\sqrt{10}\times\sqrt{10}$) and t--J ($4\times 4$) clusters 
that in the nearly half-filled case, in the presence of a strong 
onsite Coulomb repulsion and short range antiferromagnetic spin correlations, 
there were enhanced polaronic self-localization effects.
In a recent paper A. Dobry et al.\cite{dobry} 
investigated the effects
of a single dynamic $\bf q_0=(\pi,\pi)$ phonon mode coupled to 
a strongly correlated two-dimensionnal
(2D) electron system modeled by a t--J Hamiltonian. 
For cuprates superconductors, 
this $(\pi,\pi)$ mode corresponds to a planar breathing phonon
involving correlated displacements of the in-plane oxygens.
For $\sqrt{10}\times\sqrt{10}$ clusters at one-quarter filling 
A. Dobry et al.\cite{dobry} found evidence for a 
Charge Density Wave (CDW) instability at moderate values of the 
electron-phonon coupling. In addition, lattice deformations and 
CDW instabilities have been found in the one dimensional 
Holstein\cite{jorge} and t--J-Holstein\cite{fehrenbacher} models.

While both the Born-Oppenheimer and the dynamic 
single-mode approximations provide insight into this problem, one would like
to examine the dynamic multi-phonon mode system. In particular, although 
the truncation of the Hilbert space of the phonon
mode is well controlled, the single mode approach of Ref. \cite{dobry}
nevertheless assumes that in-plane oxygens oscillate in phase. 
However, there is experimental evidence that these modes 
have a small dispersion in the Brillouin zone, both in 
$La_{2-x}Sr_xCuO_4$ or in bilayer $YBa_2Cu_3O_{6+x}$ 
materials\cite{disp_LaCuO,disp_YBCO}. 
In other words, the oxygen vibrational modes are very localized and can 
be considered as independent. In the following, we investigate the influence
of taking into account a {\it finite} number M of oxygen modes by  
performing exact diagonalisation studies of finite clusters.
First, a simple model involving
the coupling of a t--J Hamiltonian with two independent $\bf q_0$
breathing modes related to different sets of in-plane oxygen atoms
is considered. Various signatures of a tendency towards
hole localization (which could lead to CDW or domain wall structures)
and the enhancement of short-range antiferromagnetic correlations between 
the remaining spins are found. 
We reach similar conclusions on a $2\times 2$ lattice when 
all the in-plane oxygen vibration degrees of freedom are taken into
account. 

We start with a generalisation of the t--J-Holstein 
Hamiltonian\cite{Heff_3bands},
\begin{eqnarray}
\label{hami}
&H& = - t \sum_{<\bf i,\bf j>, \sigma}
            ( \tilde{c}_{\bf j,\sigma}^\dagger \tilde{c}_{\bf i,\sigma}
            + \tilde{c}_{\bf i,\sigma}^\dagger \tilde{c}_{\bf j,\sigma} )
   + J \sum_{<\bf i,\bf j>}
           ( {\bf S}_{\bf i} \cdot {\bf S}_{\bf j}
           - \textstyle{1 \over 4} n_{\bf i} n_{\bf j} )  
\nonumber  
\\
   &+& \Omega \sum_{\bf i,\delta}
            ( \frac{p_{\bf i,\delta}^2}{2m} 
             +\frac{1}{2}m\Omega^2 x_{\bf i,\delta}^2 )
   + g \sum_{\bf i,\delta}
            x_{\bf i,\delta}\, 
(n_{\bf i}^h-n_{\bf i+\delta}^h)
\end{eqnarray}
\noindent
where $\tilde{c}_{\bf i,\sigma}^\dagger$ is the usual hole creation operator,
$n_{\bf i}$ and $n_{\bf i}^h$ are the electron and hole local densities 
respectively, m is the oxygen ion mass, $\Omega$ is the phonon 
frequency and $\delta=\bf x,\bf y$ differentiates the bonds along
the x- and y-direction respectively. 
Throughout energies are measured in unit of the hopping integral t.
Note that the electron-phonon g-term involves
the coupling of each copper hole with the displacements of the four
neighboring oxygens $x_{\bf i,\delta}$ and $x_{\bf i-\delta,\delta}$.
For the purpose of our discussion it is convenient to re-write the
electron-phonon coupling in momentum space,
\begin{eqnarray}
\label{hamq}
H_{e-ph} =  \Omega \sum_{\bf q,\delta} 
            (&& b_{\bf q,\delta}^\dagger b_{\bf q,\delta} + \frac{1}{2}) \\
+ \frac{1}{\sqrt{2N}}\sum_{\bf q,\delta} 
\sum_{\bf k,\sigma}\lambda_{\bf q,\delta} 
( b_{\bf q,\delta}^\dagger &&+ b_{\bf -q,\delta} )
\tilde{c}_{\bf k-\bf q,\sigma}^\dagger \tilde{c}_{\bf k,\sigma}
\nonumber
\end{eqnarray}
\noindent
where N is the number of sites, $b_{\bf q,\delta}^\dagger$ are the usual 
boson creation operators,
$\lambda_{\bf q,\delta}=\lambda_0 \sin{\frac{{\bf q\cdot\delta}}{2}}$ and
$\lambda_0=g\,\sqrt{\frac{4}{m\Omega}}$.
The Hamiltonian of Ref. \cite{dobry} is obtained from Eq. (\ref{hamq})
by (i) keeping only a single $\delta$, $\bf q=\bf q_0$ mode and
(ii) replacing the sum 
$\frac{1}{\sqrt{2N}}\sum_{\bf q,\delta}$ by unity. This is {\it a priori} 
justified by the fact that the coupling constant $\lambda_{\bf q,\delta}$
is maximum at wave vector $\bf q_0$. However, 
assuming in-phase oxygen displacements  
might overestimate cooperative effects of the phonons.
On the other hand, a complete treatment of hamiltonian (2) requires
to take the M=2N oxygen phonons into account.

The first step to go beyond the single mode approximation (SMA)
of Hamiltonian (\ref{hamq}) is to consider two independent 
bosons $b_{\bf q_0,x}$ and $b_{\bf q_0,y}$ (M=2)
for the horizontal ($\delta=\bf x$) and vertical 
bonds ($\delta=\bf y$). In this case, the sum 
$\frac{1}{\sqrt{2N}}\sum_{\bf q,\delta}$ is replaced by 
$\frac{1}{2}\sum_{\delta}$ and only two $\bf q=q_0$ modes are retained.
The ground state (GS) energies $E_0(\lambda_0)$ of the single and double mode
approximations (DMA) of Hamiltonian (\ref{hamq}) are calculated on 
a $N=\sqrt{20}\times\sqrt{20}$
cluster doped with two holes by truncating the (strictly speaking infinite) 
Hilbert space of the phonon mode(s) to a finite number of 
bosonic levels $N_{ph}$.
This can be achieved, for example, by keeping only a finite number of levels 
$n_{ph}+1$ for each oscillator. In other words, 
a maximum number of $n_{ph}$ local phonons could be excited for each 
oxygen mode. 
Then the total number of degrees of freedom for the bosons becomes
$N_{ph}=(n_{ph}+1)^M$ where M is the number of modes retained
(M=1 for SMA, M=2 for DMA and M=2N for the full problem). 

The GS 
energies $E_0(\lambda_0)-E_0(0)$ for M=1 and M=2
are displayed in Figs. \ref{fig1}(a,b) in a regime ($J=0.3$) where, in the
absence of phonons,
spin fluctuations leads to a small effective attraction between the 
holes \cite{binding_tJ}.
The dependence on the number of levels $n_{ph}+1$ is
shown in Fig. \ref{fig1}(a). 
For M=1, as in the quarter filled case considered in Ref.
\cite{dobry}, the truncating procedure is well controlled and the GS energy
decreases with $N_{ph}$ and typically converges for $N_{ph}\sim 20$ or $30$.
When more than a single mode is included, it is clear from Fig. \ref{fig1}(a)
that $n_{ph}$ (and not $N_{ph}$) is the key parameter controling the 
convergence of the GS energy. We then expect that a similar number of
levels per phonon mode ($\sim 15$) is still required to fully account for 
the vibrational degrees of freedom when $M>1$. Hence, disk space limitations 
clearly restrict the study of the full hamiltonian (2) (with $M=2N$) to small 
$2\times 2$ clusters (see below). However, useful informations can still be 
obtained with $n_{ph}\sim 2$--$5$ since most physical properties are 
smooth monotonic functions of $n_{ph}$. 
Fig. \ref{fig1}(a) suggests that polaronic effects are 
stronger when the number of independent modes (M) is increased.
As seen from Fig. \ref{fig1}(b), for increasing
$\lambda_0$ both models show a crossover from an adiabatic regime (the
energy depends weakly on the coupling strength) to a polaronic regime.
We, indeed, expect that, for a sufficiently large value of the electron-phonon
coupling, the system will start to form 
large hole polarons. By comparing the set of data for the two models 
we observe that the 
onset $\lambda_{co}$ corresponding to polaron formation 
might not depend too strongly on the truncation parameter M (or $n_{ph}$)
although for $\lambda_0>\lambda_{co}$ polaronic effects increase
with M.

In addition to the formation of polarons, the electron-phonon coupling 
(treated in the SMA)
also favors the configurations where the holes (polarons) 
stay dynamically on the same sublattice. 
Such a polaron-polaron interaction can eventually lead to a CDW instability 
as, for example, it is found at quarter filling \cite{dobry} in the SMA
or to domain walls formation. 
To complete our study, we address now the issue of the spin and charge 
correlations at low hole doping ($10\%$) and study
the influence of the number of phonon modes on the 
$(\pi,\pi)$ structure factors. 
We shall consider here only the case of commensurate $(\pi,\pi)$
charge fluctuations since the density is too close to half filling
and the system probably not large enough to accomodate any incommensurate 
order. The charge and spin structure factors at momemtum $(\pi,\pi)$
are given by 
\begin{eqnarray}
\label{st_factor}
S_c({\bf q_0})=\big<\Psi_0 | 
(\sum_{\bf i} (-1)^{(i_x+i_y)} n^h_{\bf i})^2 | \Psi_0\big>  \\
S_s({\bf q_0})=\big<\Psi_0 | 
(\sum_{\bf i} (-1)^{(i_x+i_y)} S^z_{\bf i})^2 | \Psi_0\big> 
\end{eqnarray}
\noindent
where $| \Psi_0\big>$ is the GS wavefunction calculated both in the SMA 
and DMA cases. 
$S_c({\bf q_0})$ is shown in Fig. \ref{fig2} (a) as a function of the 
electron-phonon coupling. For a single mode (SMA) a significant increase 
of the charge structure factor occurs around $\lambda_0=0.45$ indicating
a tendancy towards CDW formation or at least towards some localisation
of the charge carriers. Interestingly enough, 
Fig. \ref{fig2} (b) shows that this transition gives rise to a 
simultaneous increase of the spin-spin antiferromagnetic 
correlations. This phenomenon 
might be related to a sudden increase of the polaronic mass or to the 
localisation of the holes in some kind of (fluctuating) CDW state
or stripe (domain wall) phase. Less mobile holes clearly
lead to a weaker perturbation of the antiferromagnetic background,
hence leading to enhanced short range spin-spin antiferromagnetic correlations.
It should be noticed that, on the contrary, for quarter filling \cite{dobry} 
the CDW transition is followed by a decrease of the spin correlations. 
When two oxygen vibrational modes are considered (M=2), 
Fig. \ref{fig2} (a) and (b)
show that the increases of the charge and spin structure factors 
(indicating some hole localization) occur for approximately the 
same couplings than for M=1.

We finish our study by the full treatment of all the in-plane oxygen degrees of
freedom as seen in Fig. \ref{fig3}(a,b). Although the system size is smaller
in this case ($2\times 2$) the results are qualitatively similar. 
Fig. \ref{fig3}(b) clearly shows a large increase of the spin structure 
factor at $\lambda_{co}$. This increase becomes sharper and sharper 
when more phononic levels are included (eventhough the GS energy is not
fully converged, see Fig. \ref{fig3}(a)). Note that the cross-over value
seems to be smaller for n=0.75 ($\lambda_{co}\sim 0.3$) than 
for n=0.9 ($\lambda_{co}\sim 0.4$).

The previous results can be physically interpreted as follows;
In the SMA the oxygen atoms are constrained to oscillate in phase
leading to a dynamical potential which favors the occupation by the
carriers of one of the two sublattices. As discussed in Ref. \cite{dobry}
this can lead to a $\bf q_0$ CDW instability as can be seen for example
in a mean-field treatment. The DMA is the simplest way to introduce some
dephasing between the harmonic oscillators. 
In the extreme case of 
out-of-phase motions of the two oxygen sets the effective potential 
for the holes exactly vanishes. Therefore, one would naively 
expect that the DMA or the complete treatment of hamiltonian (2) 
lead to a weaker tendency towards CDW than the SMA.
On the contrary, our results show that, in these more reliable treatments,
the electron-phonon coupling still leads to strong self-localization 
effects \cite{tJ_Holstein}. 
More importantly, they also suggest that dynamic in-plane phonons 
could help to stabilize inhomogeneous phases \cite{Kivelson,DP_Rice}
where holes are squeezed into stripes or clumps and where
the remaining antiferromagnetic regions have enhanced 
$S_s(\pi,\pi)$ correlations. 

D. P. thanks G. Zawatsky for insightful discussions on 
experimental aspects and J. Riera for interesting correspondance.
{\it Laboratoire de Physique Quantique, Toulouse} is 
{\it Unit\'e de Recherche Associ\'e au CNRS No 505}. 
DP and DJS acknowledge support from the EEC Human Capital and Mobility 
program under Grant No. CHRX-CT93-0332 and the National Science Foundation 
under Grant No. DMR92-25027 respectively.  
We also thank IDRIS, Orsay (France), 
for allocation of CPU time on the C94 and C98 CRAY supercomputers.

%
%
\begin{figure}
\caption{Pair energy calculated on a
$\protect\sqrt{20}\times\protect\sqrt{20}$ t--J 
cluster with 2 holes for $J=0.3$ and $\Omega=0.2$.
Open (solid) symbols correspond to a single (double) $(\pi,\pi)$
breathing mode. 
(a) Energy vs number of phononic levels included in the calculations.
(b) Energy vs coupling strength $\lambda_0$. 
The number of phononic levels considered 
is indicated on the plot for each case (single and double modes).
\label{fig1}
}
\end{figure}

%
%
\begin{figure}
\caption{Charge (a) and spin (b) structure factors at momentum
$(\pi,\pi)$ vs electron-phonon coupling calculated on a 
$\protect\sqrt{20}\times\protect\sqrt{20}$ t--J 
cluster doped with 2 holes for $J=0.3$ and $\Omega=0.2$.
Open (solid) symbols correspond to a single (double) $(\pi,\pi)$
breathing mode. 
$N_{ph}=25$ phononic levels were kept in the calculation. 
\label{fig2}
}
\end{figure}

%
%
\begin{figure}
\caption{
GS energy vs $n_{ph}$ (a) and spin structure factor at momentum
$(\pi,\pi)$ vs electron-phonon coupling (b)calculated on a 
$2\times 2$ t--J 
cluster doped with 1 hole for $J=0.3$ and $\Omega=0.2$.
The total 8 independant oxygen modes have been considered.
\label{fig3}
}
\end{figure}

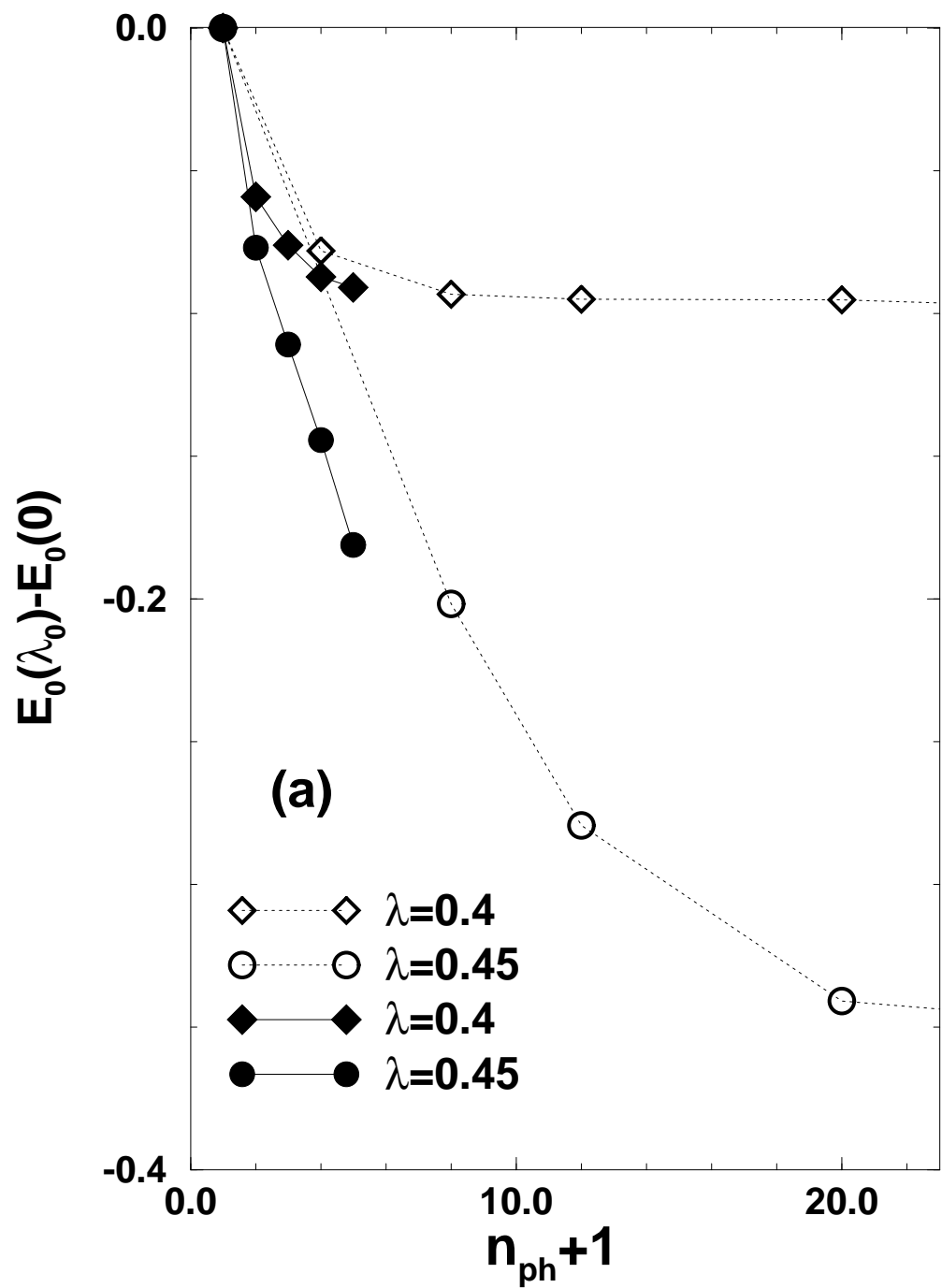
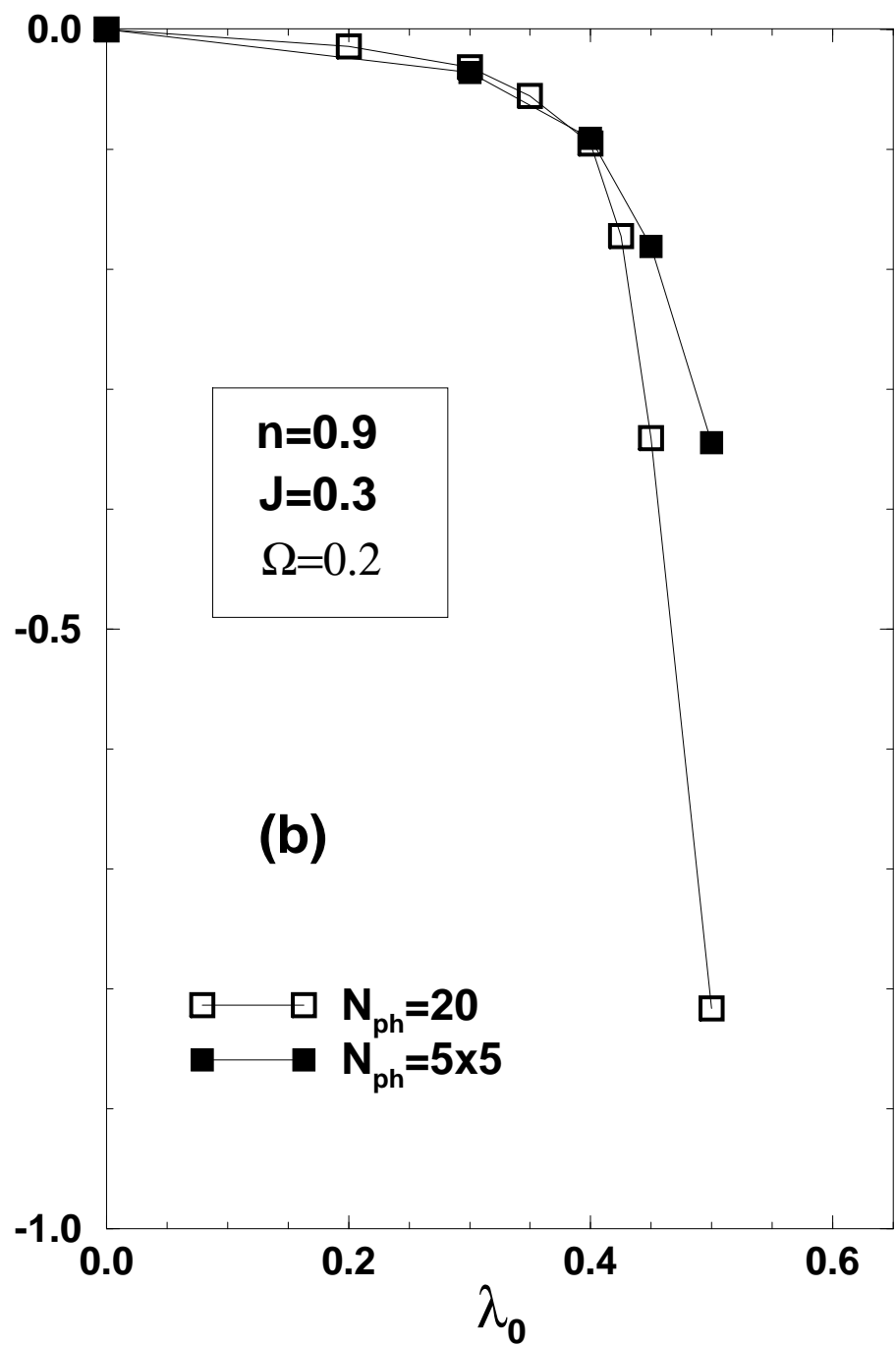
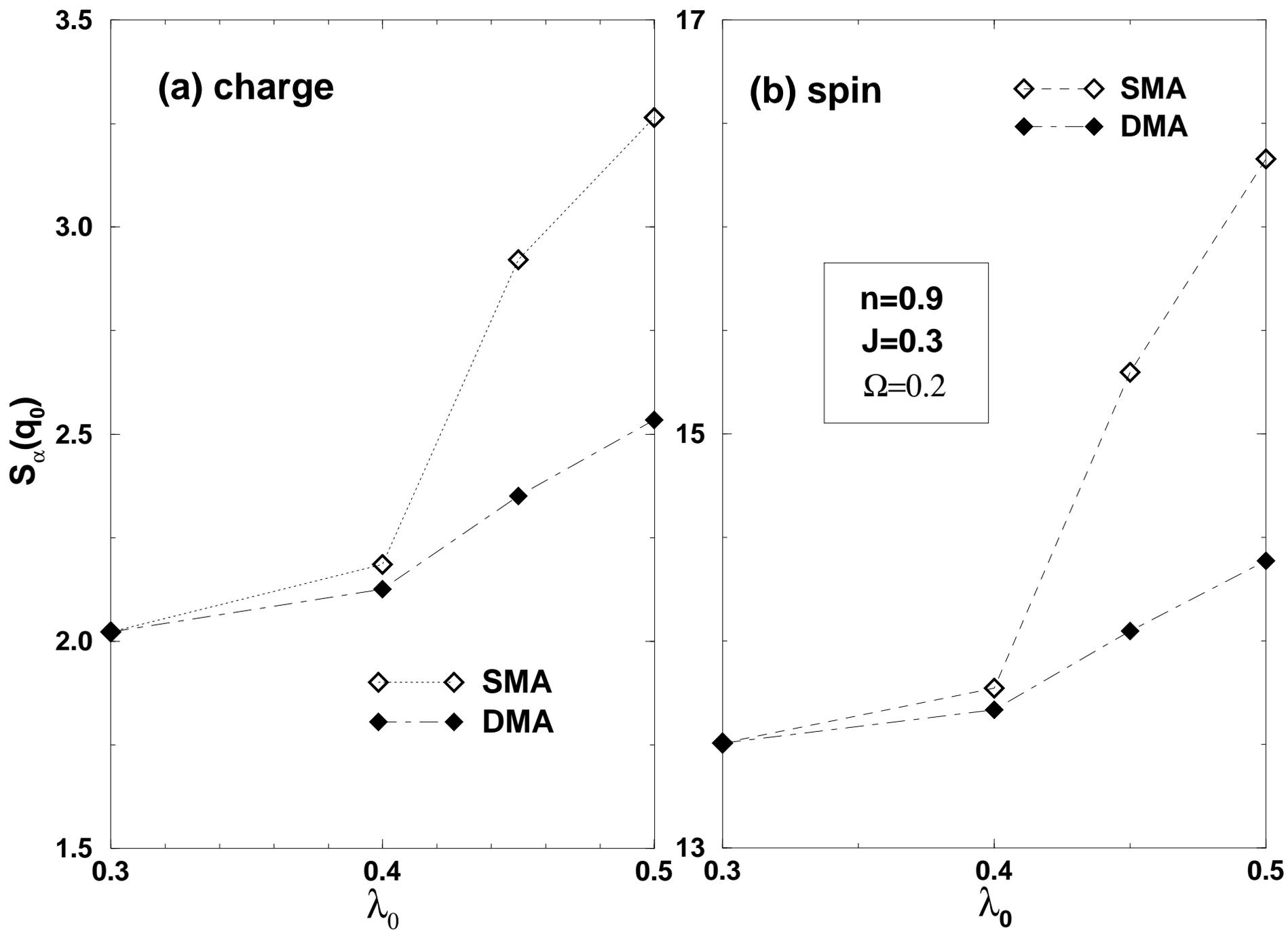
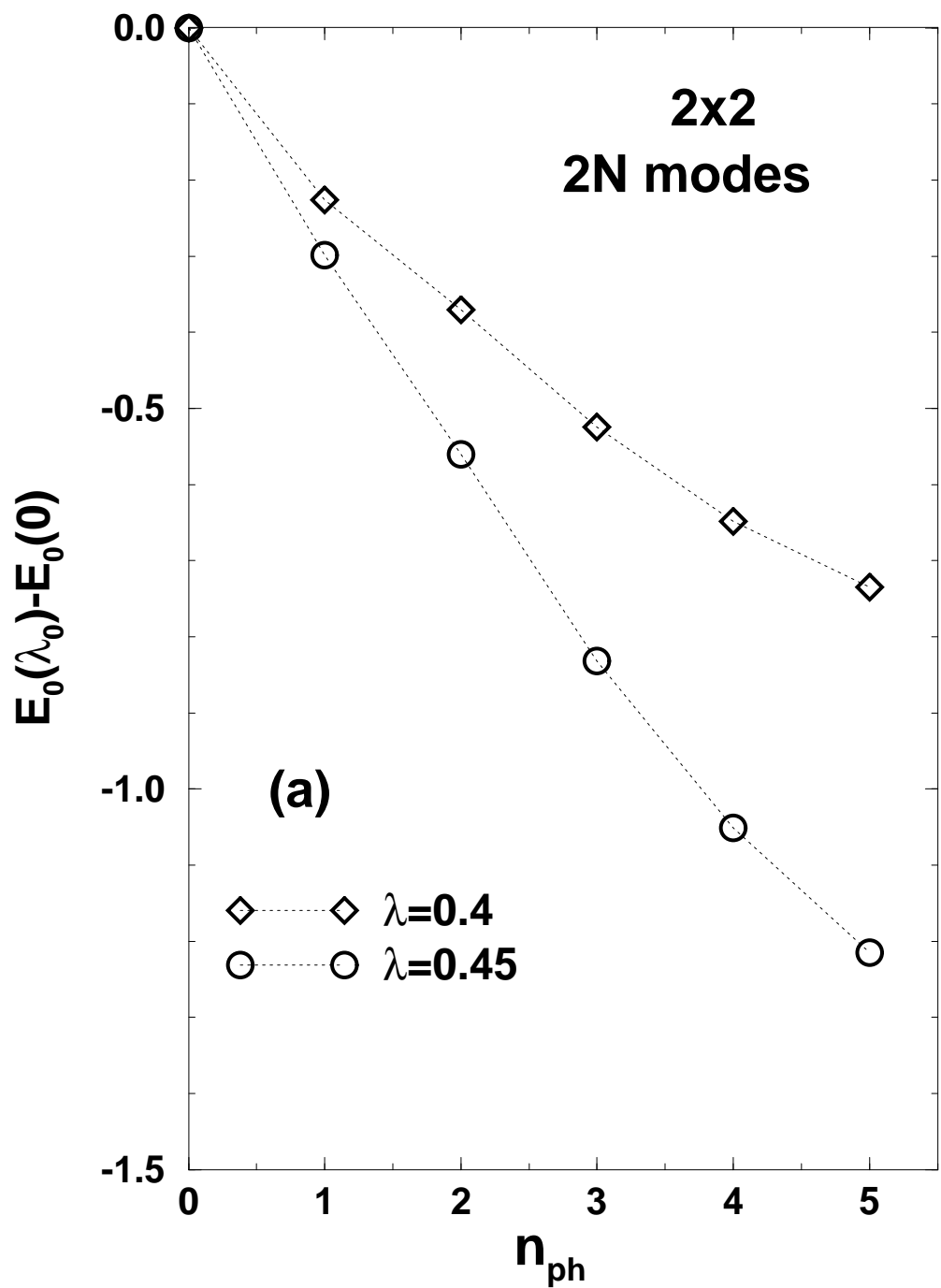
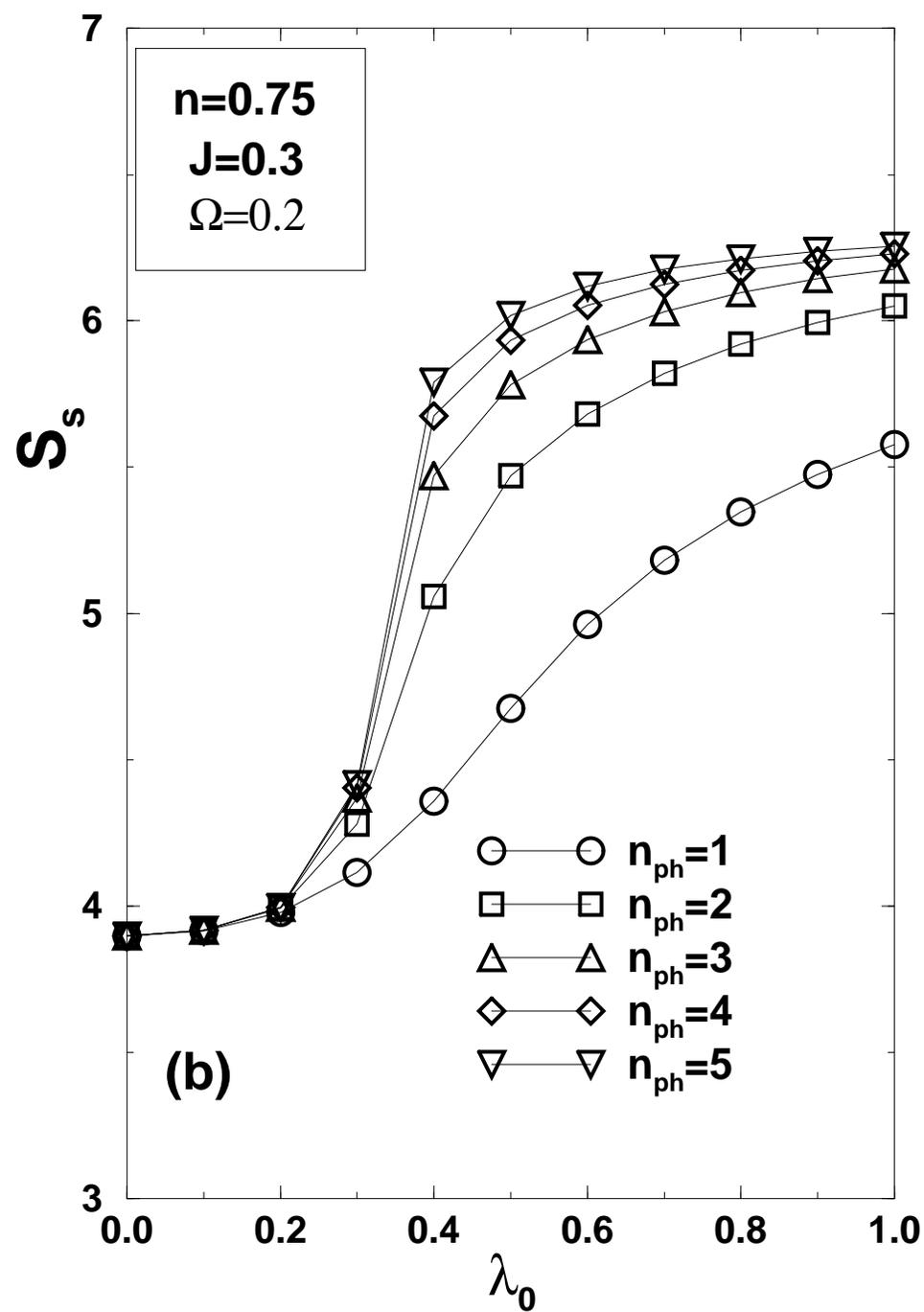
\end{document}